# Anomalously Sharp Superconducting Transitions in Overdoped $La_{2-x}Sr_xCuO_4$ Films

T.R. Lemberger,[1,*] I. Hetel,[1] A. Tsukada,[2] and M. Naito[2]

[1] *Department of Physics, The Ohio State University, Columbus, Ohio, 43210, USA*
[2] *Tokyo University of Agriculture and Technology, Tokyo, Japan*

We present measurements of *ab*-plane resistivity $\rho_{ab}(T)$ and superfluid density [$n_s \propto \lambda^{-2}$, $\lambda$ = magnetic penetration depth] in $La_{2-x}Sr_xCuO_4$ films. As Sr concentration *x* exceeds about 0.22, the superconducting transition sharpens dramatically, becoming as narrow as 200 mK near the super-to-normal metal quantum critical point. At the same time, $\lambda^{-2}(0)$ and transition temperature $T_c$ decrease, and upward curvature develops in $\lambda^{-2}(T)$. Given the sharp transitions, we interpret these results in the context of a *homogeneous d*-wave superconducting state, not a phase-separated state, with elastic scattering that is enhanced relative to underdoped LSCO due to weaker electron correlations.

*Corresponding Author: Thomas R. Lemberger, Lemberger.1@osu.edu, (614) 292-7799, (614) 292-7557 (FAX).

## I. INTRODUCTION

Inhomogeneity is a hallmark of cuprate superconductors. Beyond the considerable difficulty of preparing chemically and structurally homogeneous samples, even nominally homogeneous samples reveal electronic inhomogeneity on a several-nanometer length scale.[1] In underdoped cuprates, theory predicts spontaneous formation of stripes and possibly other short length scale structures,[2] and such structures have been observed,[3,4] so it seems that inhomogeneity is unavoidable. Overdoped cuprates are simpler because the pseudogap is absent,[5,6] and stripes are not expected. Still, several measurements on the best-studied overdoped compounds, $La_{2-x}Sr_xCuO_4$ (LSCO) and $T\ell_2Ba_2CuO_6$ (Tℓ2201), point to a superconducting state, and maybe a normal state, too, that is spontaneously inhomogeneous on an as-yet unknown microscopic length scale.[7-15]

Given the foregoing, it is remarkable that we observe sharp thermal superconductor-to-normal transitions in overdoped LSCO films, as narrow as 200 mK at a doping near the quantum super-to-normal transition. Although an inhomogeneous phase-separated state could have narrow transitions, their appearance motivates us to examine our measurements of the magnitude and T-dependence of superfluid density in the framework of a homogeneous overdoped superconducting state, with properties derived from experimental results for band structure and scattering. (Here, "superfluid density" refers to the inverse magnetic penetration depth squared, $\lambda^{-2}$, which is proportional to the imaginary conductivity, $\sigma_2$, as discussed below.) A key idea in the analysis is that electron correlations weaken with overdoping, thereby amplifying the impact of elastic scattering upon superconducting properties.

A fair bit is known about the basic properties of overdoped LSCO. Pseudogap physics is absent for $x > 0.19$.[5,6] Superfluid density at T = 0 peaks at $x \approx 0.19$ [6] and decreases as doping increases further, (more rapidly in films[16] than in bulk[17]). For the doping range of primary interest here, $x > 0.22$, the low-T crystal structure of bulk material is tetragonal - the transition from high-temperature-tetragonal (HTT) to low-temperature-orthorhombic (LTO) occurs only for $x < 0.22$.[18-20] In fact, LSCO films grown on a cubic substrate are likely to be tetragonal at any doping,[16] thereby simplifying affairs. The difficulty of controlling oxygen defects[21] during film growth means that nominally identical films may have different hole doping, but that is not an impediment. $T_c$ and resistivity serve as good secondary indicators.

Several published measurements find that the normal state of overdoped LSCO is close to that of a Fermi liquid. When $x$ exceeds about 0.22, the *c*-axis resistivity $\rho_c(T)$ changes from insulating to metallic,[20,22] as does the *ab*-plane resistivity.[23] *c vs. ab* resistivity anisotropy becomes about 100 and independent of T, the latter being characteristic of a Fermi-liquid. Importantly, *x*-ray absorption measurements indicate that the strong electron correlations present in underdoped cuprates weaken rapidly with overdoping.[24] This finding is supported by transport and heat capacity measurements[25] that also suggest a moderately correlated overdoped normal state. Angle-Resolved Photoemission Spectroscopy (ARPES) reveals a one-band Fermi surface that satisfies Luttinger's theorem and, near $x = 0.22$, crosses continuously from hole-like and centered at $(\pi,\pi)$ to electron-like and centered at (0,0).[25,26] ARPES also shows that the effective mass of electrons at the Fermi surface is independent of doping over the entire doping range where superconductivity exists, and that the Fermi velocity is four times larger along the gap-node direction, $(\pi,\pi)$, than at (1,0). All of these results will be important to our analysis of superfluid density data.

As for scattering, Narduzzo et al.[27] successfully fitted Hall coefficient, resistivity and magnetoresistance of overdoped LSCO crystals by using the Fermi surface found by ARPES, augmented with anisotropic, temperature-independent elastic scattering and isotropic, $T^2$, electron-electron scattering. This success was nontrivial because anisotropic elastic scattering was required to explain why the Hall coefficient does not change sign at the doping where the Fermi surface changes from hole-like to electron-like. For quantitative reference, they found that an overdoped, nonsuperconducting $La_{1.7}Sr_{0.3}CuO_4$ film with a residual resistivity of 13 μΩ cm had an anisotropic elastic scattering rate ranging from 80 K along the nodal direction (π,π) to 270 K along the antinodal direction (1,0). A high scattering rate seems reasonable considering that LSCO is an alloy. Our films have similar resistivities to the ones just mentioned, so we can expect the elastic scattering rate in our films to be much larger than $k_BT_c/\hbar$ everywhere on the Fermi surface.

The foregoing results point to an overdoped state that involves *d*-wave superconductivity that emerges from a moderately correlated Fermi-liquid that has a simple Fermi surface and strong, anisotropic, elastic scattering. Now we come to a key point. It is generally accepted that the underdoped *d*-wave superconducting state is "protected" from the effects of elastic scattering due to strong correlations,[29] and that the protection therefore diminishes as correlations weaken. Thus, it is reasonable to suppose that the decrease in superfluid density with overdoping is due to the enhanced effect of scattering on superconductivity, even if disorder changes little.

As mentioned above, several experiments indicate a mixed-phase superconducting state. To account for the decrease in superfluid density in overdoped cuprates, Uemura and collaborators conjectured that the overdoped state spontaneously separates into hole-poor superconducting regions coexisting with hole-rich normal metal regions, where the relative fraction of superconducting phase may increase as T decreases.[8-10] Magnetic susceptibility measurements find that when *x* exceeds 0.22, a Curie component appears and grows as *x* increases.[12] The authors suggest that the superconducting state consists of magnetic normal regions and paramagnetic superconducting regions (*i.e.*, would be paramagnetic if normal). This interpretation is supported by polarized neutron scattering results presented in the same paper. Measurements of the electronic specific heat of the overdoped superconducting state as a function of T and H were interpreted in terms of a moderately disordered normal state (elastic scattering rate of about 60 K), and an inhomogeneous phase-separated superconducting state.[14]

The following sections describe our methods for making and measuring overdoped LSCO films, present our resistivity and superfluid density data, and discuss the data in terms of a homogeneous superconducting state.

## II. EXPERIMENTAL METHODS

$La_{2-x}Sr_xCuO_4$ films are produced by molecular-beam epitaxy (MBE) on the square (001) surface of tetragonal $LaSrAlO_4$ (LSAO) substrates.[30-33] RHEED patterns show excellent layer-by-layer growth. The films' *c*-axes are perpendicular to the substrate. Compressive strain due to the 0.6% mismatch between film and substrate ($a$ = 3.754 Å for $LaSrAlO_4$ and 3.777 Å for bulk LSCO) gives our films a maximum $T_c$ (44 K) that is above the maximum $T_c$ in LSCO crystals.[31,33] The maximum $T_c$ occurs at the same doping, $x \approx 0.15$, for films and bulk. At the same that compression increases $T_c$, it lowers the *ab*-plane resistivity.[34] The mechanism underlying the enhanced superconductivity is under investigation.[33] For reference, we note Bozovic *et al*. have made a detailed study of the microstructure of MBE-grown LSCO films on LSAO.[35] Sr doping values are nominal. They are set by atomic beam fluxes during deposition. Sample properties are listed in Table I.

After the first series of films (thickness $d$ = 45 nm) was grown, noting the jump in properties between $x$ = 0.24 and $x$ = 0.27, we decided to grow a film at $x$ = 0.30 and a second film at $x$ = 0.27 (both with $d$ = 90 nm) to get more data points near the QPT. (See shaded rows at bottom of Table I.) These films were grown with a slightly different protocol, aimed at keeping oxygen stoichiometry at 4.0, and a greater thickness since that change seemed to improve film properties somewhat. They have somewhat higher $T_c$'s and superfluid densities than for the first series, perhaps due in part to a slight difference in oxygenation.

Two samples were grown simultaneously at each Sr concentration, one on a narrow substrate for measuring resistivity and the other on a 10×10×0.35 $mm^3$ substrate for measuring superfluid density. *ab*-plane resistivities $\rho_{ab}(T)$, Fig. 1, are obtained from standard four-point measurements.

Low frequency sheet conductivity, $\sigma d = \sigma_1 d - i\sigma_2 d$, is measured with a two-coil mutual inductance technique, with drive and pickup coils on opposite sides of the film.[36,37] Coil dimensions are about 2 mm. A low-frequency (50 kHz) current in the drive coil produces a small *ac* magnetic field that is attenuated by eddy currents induced in the sample. The attenuation is approximately proportional to the magnitude of $\sigma d$. Care is taken to ensure that the *ac* field is small enough that measurements are taken in the linear response regime. The conductivity, $\sigma$, is obtained by dividing $\sigma d$ by film thickness $d$. The superfluid density is defined from the nondissipative part of $\sigma$: $\lambda^{-2} \equiv \mu_0\omega\sigma_2$.

Table I. Properties of overdoped $La_{2-x}Sr_xCuO_4$ films grown by MBE on LSAO (100) substrates. Sr concentrations are nominal. $\Delta T_c$ is the width of the peak in $\sigma_1$ near $T_c$. The last two films (shaded) were grown later than the others, with a slightly different protocol, and are twice as thick as the others.

| $x$ | $T_c(\rho = 0)$ (K) | $T_c(\lambda^{-2}=0)$ (K) | $\lambda^{-2}(0)$ ($\mu m^{-2}$) | $\rho_{ab}$ (50 K) ($\mu\Omega$ cm) | $\Delta T_c$ (K) |
|---|---|---|---|---|---|
| 0.15 | 44 | 42 | 17.4 | 90 | 4 |
| 0.18 | 41 | 38 | 21.5 | 54 | 6 |
| 0.21 | 33 | 32. | 20.3 | 48 | 4 |
| 0.24 | 19 | 18.5 | 11.1 | 37 | 3 |
| 0.27 | 4.0 | 3.9 | 0.15 | 31 | 2 |
| 0.27 | 21 | 20 | 6.8 | 70 | 1.6 |
| 0.30 | 9.0 | 8.5 | 1.6 | 56 | 0.2 |

## III. RESISTIVITY

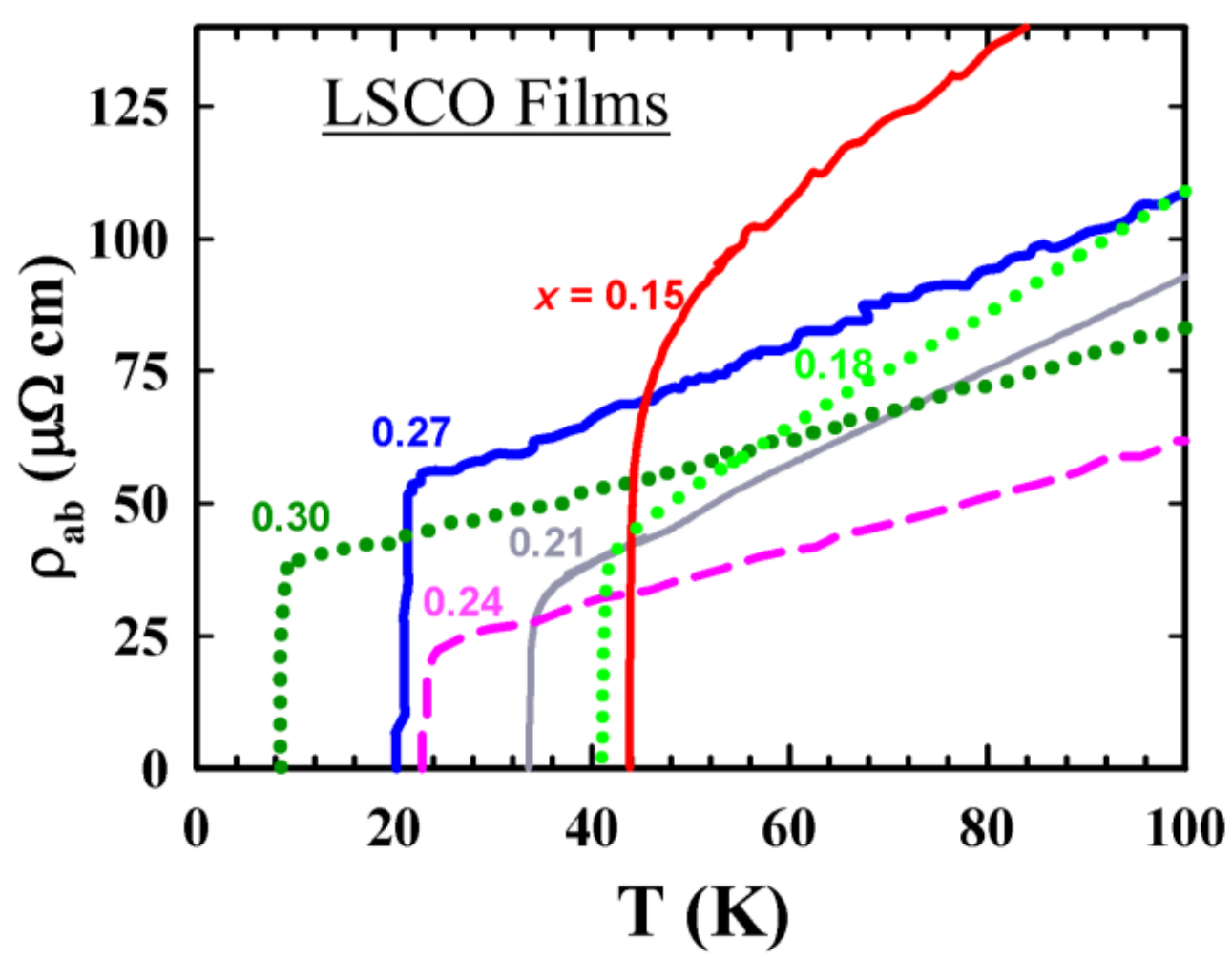

Fig. 1. (Color online) *ab*-plane resistivity $\rho_{ab}(T)$ below 100 K for overdoped $La_{2-x}Sr_xCuO_4$ films.

*ab*-plane resistivities of our films have a minimum at x ≈ 0.24, Fig. 1, achieving a low residual resistivity (about 40 μΩ cm) at *x* = 0.30, comparable to that of a similarly overdoped

LSCO crystal.[28] For comparison, films grown on $SrTiO_3$ substrates are under slight tension, and they have a minimum in their residual resistivity of about 40 μΩ cm at $x \approx 0.30$.[23] When examined closely, resistivity shows a subtle change in behavior. For $x$ less than about 0.22, resistivities extrapolate to near zero at T ≈ 0. For larger $x$, resistivity above $T_c$ is flatter, and it extrapolates to a nonzero value at T = 0. Finely spaced dopings are needed to explore this region in detail.

We define the resistive $T_c$ from where $\rho_{ab}$ vanishes, Table I. This agrees with $T_c$ defined from where superfluid appears, as discussed below. $T_c$ evolves smoothly with doping, reaching a maximum of 44 K at $x \approx 0.15$. Resistive transitions are sharp at all Sr concentrations, but resistivity is an unreliable measure of transition width because it probes only the first superconducting path through the sample. The width of the "fluctuation" peak in $\sigma_1(T)$ is better, as discussed in Sec. V.

## IV. REAL CONDUCTIVITY $\sigma_1$(T) AND SUPERFLUID DENSITY $\lambda^{-2}$(T).

Figures 2 and 3 show $\lambda^{-2}(T)$ and $\sigma_1(T)$ for slightly- and strongly-overdoped films, respectively. The value of $\lambda^{-2}(0)$ near optimal doping ($x \approx 0.15$) is comparable to that of LSCO powders,[6] (and generally a bit larger that has been reported for other bulk[6] and film[16] samples), again indicating good film quality. $\lambda^{-2}(0)$ decreases with overdoping, as has been observed in other cuprates.[6] The dashed blue curves in Fig. 2 are quadratic fits to $\lambda^{-2}(T)$ at low T, showing that films near optimal doping are consistent with the low-T quadratic behavior expected for disordered *d*-wave superconductors.[39] Also, they show that the evolution toward upward curvature at intermediate temperatures is present already at optimal doping.

Over the small doping change from $x = 0.21$ to 0.24, big changes occur. $T_c$ and $\lambda^{-2}(0)$ both decrease by about 40% (from 33 K to 20 K and 20 to 11.5 $\mu m^{-2}$, respectively), while the low-T quadratic behavior is either restricted to T < 2 K, or it is replaced by T-linear. The slope of $\lambda^{-2}(T)/\lambda^{-2}(0)$ just below $T_c$ decreases, giving rise to strong upward curvature at intermediate temperatures.

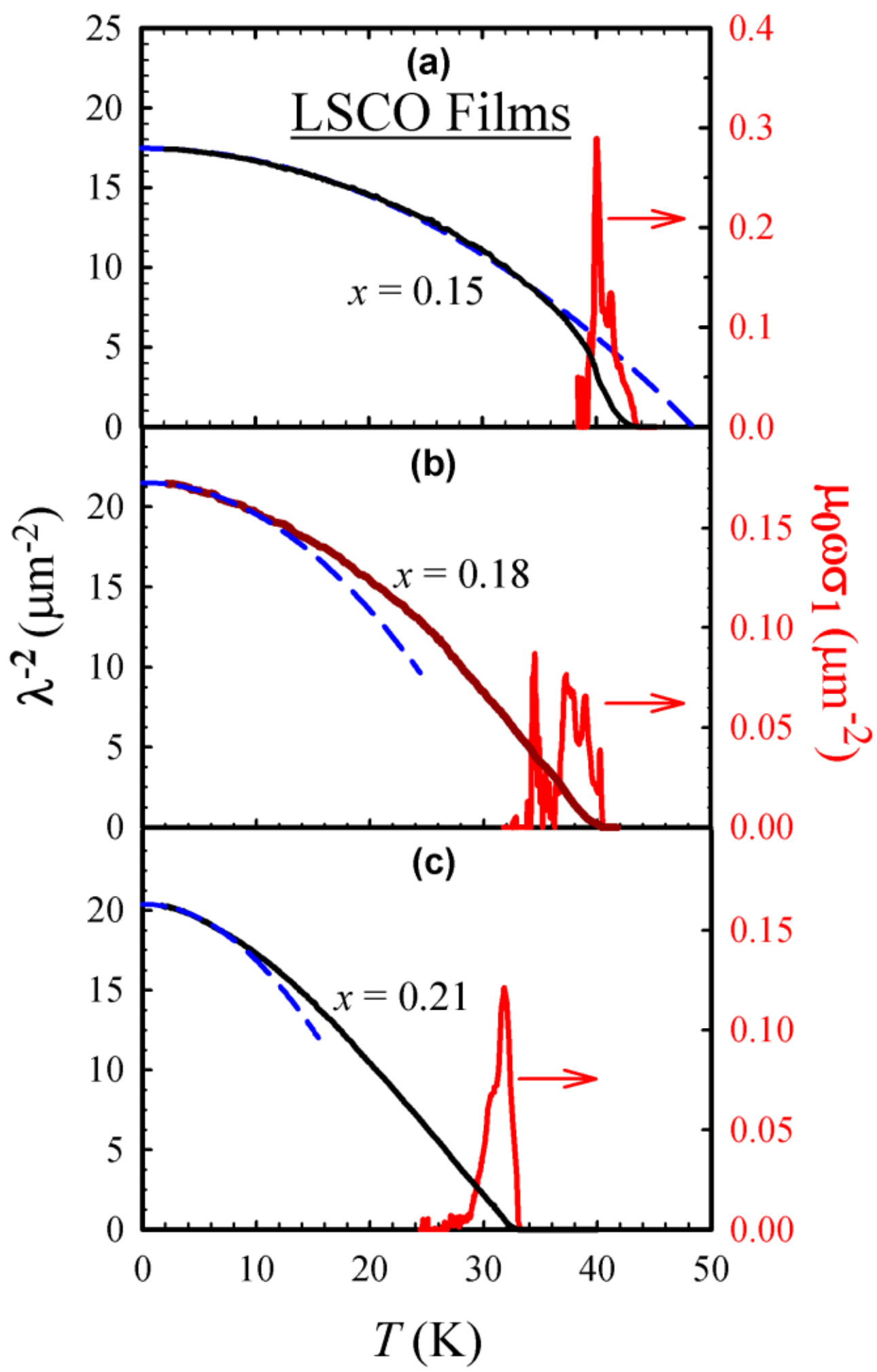

Fig. 2. (Color online) $\lambda^{-2}(T)$ (dark curves) and $\sigma_1(T)$ (red peaks) measured at 50 kHz for $La_{2-x}Sr_xCuO_4$ films near optimal doping: (a) $x$ = 0.15, (b) $x$ = 0.18, (c) $x$ = 0.21. Dashed blue curves are quadratic fits to the low-T data.

The peaks in $\sigma_1$ (Figs. 2 and 3) near $T_c$ can be good indicators of sample homogeneity. Films with $x \leq 0.21$, (including underdoped films not shown here[40]), have peaks several K wide that show structure indicative of multiple $T_c$'s, either in different layers in the film or laterally over the mm-scale area probed by our measurements. By contrast, films with $x \geq 0.24$ (*i.e.*, with $T_c/T_c^{max} \leq ½$) show structureless peaks about 1 K wide. A remarkably sharp 200 mK transition width is achieved in a film very close to the super-to-normal metal quantum phase transition, *i.e.*, $T_c$ is less than 10% of $T_c^{max}$, and superfluid density $\lambda^{-2}(0)$ less than 1% of the maximum value at $x \approx 0.18$. The intersection of the KTB line with $\lambda^{-2}(T)$ in Fig. 3b indicates where a 2D transition is predicted, assuming that the film fluctuates as a single 2D entity rather than independent layers. Note that the predicted downturn in $\lambda^{-2}(T)$ appears at this point. The significance of this feature is discussed below.

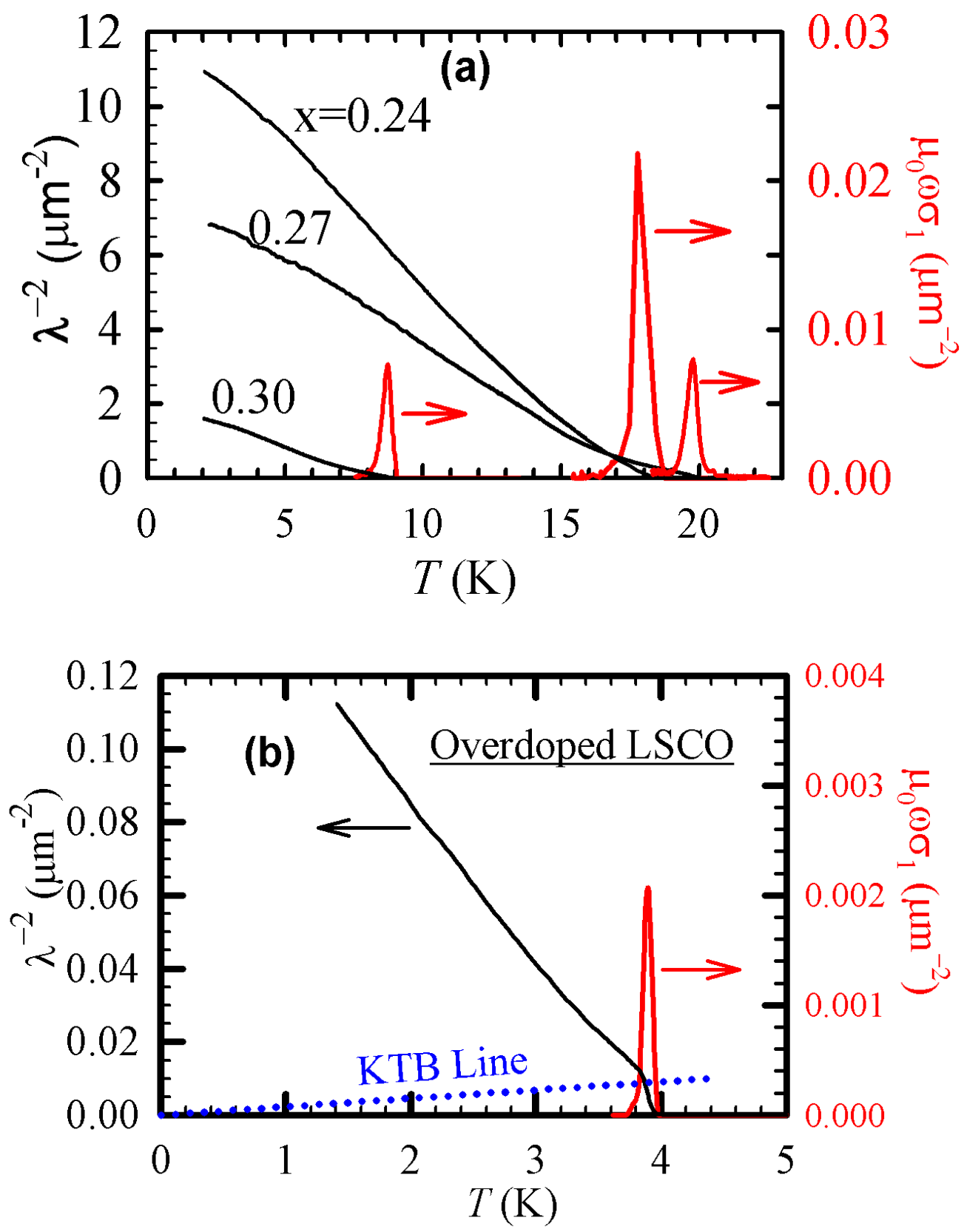

Fig. 3. (Color online) $\lambda^{-2}$(T) (dark curves) and $\sigma_1$ (red peaks) measured at 50 kHz for strongly overdoped $La_{2-x}Sr_xCuO_4$ films with nominal dopings: (a) *x*=0.24, 0.27 and 0.30; (b) a second, thicker film with *x*=0.27, with the KTB line discussed in the text.

## V. DISCUSSION

In this section, we first discuss the meaning of the peak in $\sigma_1$. Next, we discuss how the superfluid density $\lambda^{-2}(0)$ can decrease with overdoping in the framework of a homogeneous disordered *d*-wave superconductor. The key idea is that electron correlations weaken with overdoping, thereby increasing the effect of disorder on superconductivity. We then interpret the evolution of the T-dependence of $\lambda^{-2}(T)$ with overdoping in the context of the Fermi surface obtained from ARPES and anisotropic elastic scattering deduced from normal-state transport measurements. Having accounted for the main features of our data on LSCO in a mean-field framework, we discuss why it may be reasonable to neglect the effects of thermal phase fluctuations. Finally, recalling the evidence for spontaneous phase separation in overdoped LSCO, we note that the over-electron-doped cuprates behave much like over-hole-doped LSCO, even though there is little evidence for phase separation in them.

We begin with the "fluctuation" peak in $\sigma_1$ near $T_c$. Phenomenologically, the peak arises from the crossover in low-frequency film impedance from resistive above $T_c$ to mainly inductive below $T_c$. It is associated with thermal vortex fluctuations in the sense that these fluctuations mediate a continuous decrease of resistance to zero, rather than a discontinuous jump. The width of the peak in $\sigma_1$ is related to how rapidly resistance decreases, and it is *nonzero even for a perfectly homogeneous superconductor*. If there is a single peak, then its width is an *upper limit* on the spread of $T_c$'s in the sample. Inhomogeneity in $T_c$ generally spreads the decrease in resistance out over a wider temperature range, thereby broadening the peak in $\sigma_1$. When different layers in a film have distinctly different $T_c$'s, by accident or by construction, a peak is observed at each $T_c$, which broadens the peak in $\sigma_1$ and causes substructure rather than a single clean peak.

The clean narrow transitions of strongly overdoped LSCO films are striking when contrasted with those of moderately overdoped films, Fig. 2, made in the same system under essentially the same conditions, and therefore likely to possess similar degrees of structural and chemical homogeneity. As seen in Fig. 2, moderately overdoped films have peaks with structure indicative of several closely spaced $T_c$'s, whereas overdoped films do not. It looks as though there is a transition of some sort at $x \approx 0.22$, especially in light of the changes in crystal structure,[18-20] Fermi surface,[25,26] and resistivity[20,22] that occur at the same doping.

An interesting quantitative context is provided by considering the effect of variations in doping. At optimal doping, the slope $dT_c/dx \approx 0$, so one might expect to find the sharpest transitions there, as one finds in electron-doped cuprate films.[41,42] Near the doping $x \approx 0.30$ where superconductivity disappears, $dT_c/dx$ should be about -6 K/.01, estimated in the usual parabolic approximation to the dependence of $T_c$ on $x$.[43] Thus, a mere 1% doping variation, *e.g.*, $0.297 \leq x \leq 0.300$, that might occur during film deposition would result in a 2 K spread of $T_c$'s through the film thickness, ten times larger than the 0.2 K wide peak in $\sigma_1$ that we observe. Thus, we believe that interlayer coupling is strong enough to homogenize $T_c$ through the film thickness, even though *c vs. ab* plane resistive anisotropy is large (≈ 100).

The abrupt sharpening of transitions for $x > 0.22$ is no doubt abetted by the improved interlayer coupling, as indicated by the transition in *c*-axis resistivity from insulating to metallic.[18,20] It would seem likely that phase separation into normal and superconducting regions would work in the opposite direction, although this is not necessarily so. In the end, as noted above, we are motivated to present a model based on a homogeneous *d*-wave superconductor that accounts successfully for the main features of our data.

First, to account for the decrease in $\lambda^{-2}(0)$ with overdoping, we look to the destructive effect of elastic scattering on *d*-wave superconductivity rather than phase separation. This seems a doubtful enterprise because disorder cannot change significantly when $x$ changes from, say, 0.21 to 0.24, and we need to account for almost factor-of-two reductions in $\lambda^{-2}(0)$ and $T_c$. But, recall that the *d*-wave superconducting state of *underdoped* cuprates is surprisingly insensitive to impurity scattering,[29,39] a phenomenon that has been explained as a byproduct of very strong electron correlations.[29] Correlations are much weaker in the overdoped state, as revealed by recent *x*-ray absorption measurements,[24] so the effect of elastic scattering should grow rapidly with overdoping, even if the underlying disorder hardly changes. Given that the normal state elastic scattering rate in overdoped LSCO ranges from $\hbar/k_B\tau \approx 80$ K to 270 K on the Fermi surface,[27] and $T_c$ is below 40 K, the effect could be dramatic.

For dirty *d*-wave superconductors, $T_c$ is proportional to $[\lambda^{-2}(0)]^{1/2}$ in the strong scattering limit.[39] In fact, this is what we see when $x \geq 0.24$.[44] There is still the possibility that superconductivity ends at a 3D quantum critical point that has square root scaling, as happens at the underdoped quantum phase transition in YBCO,[45-48] but the simpler explanation is dirty *d*-wave superconductivity.

Now we consider the T-dependence of normalized superfluid density, $\lambda^{-2}(T)/\lambda^{-2}(0)$. Moderately overdoped films, $0.15 \leq x \leq 0.21$, show the quadratic low-T behavior that is most naturally interpreted as disordered *d*-wave superconductivity.[39] More strongly overdoped films, $x \geq 0.24$, appear to lose the quadratic low-T behavior, possibly becoming T-linear. However, we believe that the right way to view the data is to focus on suppression of the slope of $\lambda^{-2}$ near $T_c$. This suppression probably pushes the low-T quadratic behavior below our lowest experimental temperature.

We argue that the upward curvature in $\lambda^{-2}(T)$ is qualitatively consistent with ARPES band structure plus scattering. Our argument is two-pronged. First, we argue that at T = 0 more superfluid comes from the nodal regions of the Fermi surface, near $(\pi,\pi)$, than from the antinodes due to the much higher Fermi velocity[25,26] and much smaller scattering rate near the nodes.[27] Second, it is known that in *d*-wave superconductors, regions of the Fermi surface "become superconducting" when thermal energy $k_BT$ drops below the local value of the superconducting gap, $\Delta(\mathbf{k},T)$. Hence, the antinodal regions "become superconducting" pretty quickly as T drops below $T_c$, while the nodal regions turn on at lower temperatures. The slope of $\lambda^{-2}(T)$ is therefore larger at low T than just below $T_c$. It is entirely possible that $\lambda^{-2}$ is quadratic in T below our lowest experimental temperature. Calculations are needed to test the validity of our proposed homogeneous superconducting state, but the qualitative idea is sound.

Since the appearance of upward curvature in $\lambda^{-2}(T)$ looks odd, we pause to note that a qualitatively similar evolution toward upward curvature in the overdoped state has been observed previously, in $\lambda^{-2}(T)$ of LSCO[49] and Tℓ2201[50] bulk powders. Indeed, some time ago Paget et al.[51] found in LSCO films, all near optimal doping, that the T-dependence of $\lambda^{-2}(T)$ seemed to have either of two distinct forms, one with much more downward curvature than the other, as if the change in shape occurred over a small doping interval.

Having argued that our results appear to be describable in terms of a homogeneous disordered *d*-wave superconducting state, we must ask whether it is reasonable to neglect thermal phase fluctuations. In fact, an order-of-magnitude calculation finds that if each $CuO_2$ layer fluctuated independently, then classical low-T thermal phase fluctuations would produce a linear suppression in superfluid density about the size that we observe at our lowest experimental temperature. Two things argue against this scenario.

First, there is evidence for significant interlayer coupling. In the most overdoped film, Fig. 3(b), the intersection of the dotted KTB line with the measured $\lambda^{-2}(T)$ marks where a 2D phase transition is predicted, assuming that the 75-layer-thick film behaves *as a single 2D entity*, not 75 independent $CuO_2$ layers. The predicted 2D drop in superfluid density is observed at the intersection, along with a corresponding peak in $\sigma_1(T)$. Moreover, microwave measurements on *underdoped* LSCO films find similar behavior,[51,52] indicating significant interlayer coupling in underdoped LSCO, too.

Second, if layers were independent, then the 2D transition would be predicted at a temperature a little below 1 K (*i.e.*, dotted KTB line would be about 75 times steeper), and we should not be able to observe superconductivity in our apparatus. There exist theoretical mechanisms whereby the transition would not appear in the superfluid density at the single-layer 2D transition temperature, even though layers are very weakly coupled.[*e.g.*, Ref. 53] However, it is difficult to see how fluctuations would produce the observed upward curvature leading up to $T_c$.

The electron-doped cuprates, *e.g.*, $La_{2-x}Ce_xCuO_4$ (LaCeCuO) and $Pr_{2-x}Ce_xCuO_4$ (PrCeCuO), provide an interesting point of comparison because their tetragonal crystal structure is close to LSCO's (at $x > 0.22$), but evidence for spontaneous phase separation is weak, although the issue is controversial.[see review in Ref. 54] Resistivities and superfluid densities of LaCeCuO and PrCeCuO films near optimal doping[41,42] are about the same as in our LSCO films. *c vs. ab* resistive anisotropy of e-doped cuprates slides from 1000 at optimal doping to 100 at the over-electron-doped quantum phase transition,[55] again similar to LSCO. LaCeCuO and PrCeCuO films have clean, narrow $\sigma_1$ peaks ($\lesssim$ 1 K), like in LSCO, although narrow transitions also occur for under-electron-doped films. The Fermi surface of over-electron-doped cuprates evolves from electron-like to a hole-like Fermi surface centered at ($\pi$,$\pi$), but details are more complicated than for LSCO.[see *e.g.*, Ref. 56 and the review in Ref. 54] Finally, overdoped LaCeCuO and PrCeCuO films develop an upward curvature in superfluid density due to a suppression in slope just below $T_c$. The upshot is that LSCO's e-doped cousins behave a lot like LSCO, even though they are not likely to experience phase separation.

## VI. CONCLUSION

When doping in $La_{2-x}Sr_xCuO_4$ exceeds $x \approx 0.22$ (*i.e.*, $T_c/T_c^{max} \lesssim ½$), interlayer coupling strengthens, the HTT to LTO transition disappears, and electron correlations weaken dramatically. These changes facilitate narrow superconducting transitions. Superfluid density, $\propto \lambda^{-2}(0)$, decreases with doping, and $\lambda^{-2}(T)$ develops upward curvature. These effects are consistent with the Fermi-liquid band structure obtained from ARPES plus the anisotropic normal-state elastic scattering obtained from transport measurements, given the enhanced effect that elastic scattering has on *d*-wave superconductivity when the strong correlations of the underdoped state weaken in the overdoped state. We acknowledge that the overdoped superconducting state may spontaneously phase separate into hole-rich and hole-poor regions, but we emphasize that phase separation is not necessary to understand the present measurements.

Finally, we comment that it is possible that bulk samples phase separate but our films do not, or not as much, because our films are under compression. However, given the similarities in terms of magnitude and T-dependence of $\lambda^{-2}$ between films and bulk, it seems unlikely that the superconducting states are so different on a microscopic length scale.

ACKNOWLEDGEMENTS: This work was supported in part by NSF DMR grant 0203739 and DOE grant FG02-08ER46533. We acknowledge numerous interesting discussions with Mohit Randeria and David Stroud.